\documentclass[aps,prb,twocolumn,superscriptaddress,showpacs,longbibliography,nofootinbib]{revtex4-1}
\usepackage{amsmath}
\usepackage{graphicx}
\usepackage{amsfonts}
\usepackage{verbatim}
\usepackage{amssymb}
\usepackage{color}
\usepackage{dsfont}
\usepackage{amsmath} % or simply amstext
\usepackage{hyperref}
\usepackage{physics}
\usepackage{soul}
\hypersetup{colorlinks = true, urlcolor = blue, linkcolor = blue, citecolor = blue}
\usepackage[normalem]{ulem}

\begin{document}
\title{Symmetric, off-diagonal, resistance from rotational symmetry breaking in graphene-WSe$_2$ heterostructure: prediction for a large magic angle in a Moire system}
\author{Jay D. Sau} 
%\author{Sumanta Tewari}
\affiliation{Department of Physics, Condensed Matter theory center and the Joint Quantum Institute, University of Maryland, College Park, MD 20742}
\author{Sumanta Tewari}
\affiliation{Department of Physics and Astronomy, Clemson University, Clemson, SC 29634}
\date{May 2023}

\begin{abstract}
    We show that any two-dimensional system with a non-zero \textit{symmetric} off-diagonal component of the resistance matrix, $R_{xy}=R_{yx} \neq 0$, must have the in-plane rotational symmetry broken down to $C_2$. Such a resistance response is Ohmic, and is different from the Hall response which is the \textit{anti-symmetric} part of the resistance tensor, $R_{xy}=-R_{yx}$, is rotationally symmetric in the 2D plane, and requires broken time-reversal symmetry. We show how a minute amount of strain due to lattice mismatch - less than $1 \%$ - can produce a vastly exaggerated symmetric off-diagonal response - $\frac{R_{xy}}{R_{xx}} \sim 20\%$ - because of the momentum matching constraints in a Moire system. Our results help explain an important new transport experiment on graphene-WSe$_2$  heterostructures, as well as are relevant for other experimental systems with rotational symmetry breaking, such as nematic systems and Kagome charge density waves. Additionally, our work predicts an example of a `magic' angle, $\theta\sim 27^0$, in a Moire system which is a significant fraction of $\pi$. Our prediction that the anomalous resistance anisotropy occurs at a large value of magic angle, in contrast to known examples of magic angle transport anomalies that are small fractions of $\pi$, can be experimentally tested in graphene-WSe$_2$ heterostructures.        
\end{abstract}

\maketitle

\textit{Introduction:} Two-dimensional graphene-based heterostructures have attracted intensive attention in recent years as a versatile platform where the substrate can qualitatively modify the graphene spectra by inducing, e.g., a spin-orbit coupling or a Moire superlattice \cite{Qiao, Dean, Cao, Wakamura, Tiwari}. Such effects from the substrate, e.g., from WSe$_2$ in the graphene-WSe$_2$ heterostructure, provide new opportunities to tailor graphene bandstructure, which may lead to the experimental realization of interesting  observable effects   theoretically predicted for graphene.~\cite{Xiao, Cresti, Mak, Lee, Liu1, Wallbank, Qiao1, Shimazaki, Sui} In an important new experiment on high-mobility graphene-WSe$_2$ heterostructure, \cite{Aveek_Bid} a symmetric off-diagonal resistance, $R_{xy}=R_{yx}$, has been found to be non-zero even in the absence of an external magnetic field. This off-diagonal resistance, which is symmetric in the spatial coordinates, has been termed a symmetric Hall effect and has been found to exist without time-reversal symmetry breaking, and persists up to room temperature. The conventional Hall resistance, on the other hand, is antisymmetric in the spatial coordinates, $R^H_{xy}=-R^H_{yx}$, and can be non-zero only when the time-reversal symmetry is broken in the system. \cite{Buttiker, Onsager1, Onsager2}  In this Letter, we show that such a symmetric transverse resistance, leading to a transverse electric voltage in response to an applied longitudinal current, can be explained in terms of rotational symmetry breaking due to lateral uniaxial strain on the graphene layer.  
%which can break the $C_3$ symmetry of the parent graphene layer down to $C_2$ or below in the heterostructure.
Most importantly, we show that even a small amount of strain (less than $\sim 1\%$) can have quite an exaggerated effect on the symmetric transverse resistance ($\frac{R_{xy}}{R_{xx}} \sim 20 \%$) in a Moire system, as has been found experimentally.~\cite{Aveek_Bid}
%, e.g., graphene-WSe$_2$ heterostructure. 

In a two-dimensional system the current $\vec{j}$ and the electric field $\vec{E}$ are related by the conductivity tensor $\sigma_{\alpha \beta}$ ($\alpha, \beta =x,y$) by the relation $j_{\alpha}=\sigma_{\alpha \beta} E_{\beta}$. The tensor $\sigma_{\alpha \beta}$ can be decomposed into a symmetric ($\sigma^s_{\alpha \beta}$) and an anti-symmetric ($\sigma^a_{\alpha \beta}$) component, satisfying $\sigma^s_{\alpha \beta}=\sigma^s_{\beta \alpha}$ and $\sigma^a_{\alpha \beta}=-\sigma^a_{\beta \alpha}$, respectively. The current can then be written as, $j_{\alpha}=\sigma^s_{\alpha \beta}E_{\beta} + \sigma^a_{\alpha \beta}E_{\beta}$. 
%Since a $2\times2$ anti-symmetric matrix $\sigma^a_{\alpha \beta}$ has only one independent element and is equivalent to a scalar, we can associate it with a vector pointing in the $\hat{z}$-direction. 
We write the antisymmetric $2\times 2$ matrix $\sigma^a_{\alpha\beta}$  as, $\sigma^a_{\alpha\beta}=\gamma B_z\epsilon_{\alpha\beta}$, where $\gamma$ is a coupling constant, $\epsilon_{\alpha\beta}$ is the 2D Levi-Civita symbol and $\vec{B}=B_z\hat{z}$ is the external magnetic field. This leads to, $j_{\alpha}=\sigma^s_{\alpha \beta}E_{\beta} - \gamma\epsilon_{\alpha \beta} B_z E_{\beta}=\sigma^s_{\alpha \beta}E_{\beta}+\gamma(\vec{E}\times (B_z\hat{z}))_{\alpha}$, where $\vec{B}$, assumed to point in the $\hat{z}$-direction, enters the expression for the antisymmetric Hall conductivity, which is a scalar with respect to rotation in the $x-y$ plane. With Joule heating being proportional to $\vec{j}\cdot\vec{E}=j_{\alpha}E_{\alpha}$, it is clear that the Joule heating is non-zero for the part $\sigma^s_{\alpha \beta}E_{\beta}E_{\alpha}$, which is thus Ohmic, and zero for the part $(\vec{E}\times \vec{B})_{\alpha}E_{\alpha}$. Since a current density $j_\alpha$ is odd under time-reversal symmetry while the electric field is even, the conductivity $\sigma$ should break time-reversal symmetry. The dissipative part $\sigma^s$ breaks time reversal through the generation of entropy in the system i.e. heating. Since the antisymmetric part $\sigma^a$ is non-dissipative and does not generate entropy, the time-reversal breaking of the conductance must be intrinsic to the system. Such time-reversal breaking in transport was discovered as the Hall conductivity in conductors in magnetic fields. Thus $\sigma^a$ must vanish in the absence of magnetic field or ferromagnetism.  
%Conventionally, the non-dissipative anti-symmetric part of the conductivity tensor is identified with Hall conductivity, while the symmetric, dissipative part is the Ohmic conductivity. 
Note that even the Ohmic part of the conductivity tensor, $\sigma^s_{\alpha \beta}$, which can be non-zero even in time-reversal invariant systems, can have a non-zero off-diagonal component, which may lead to a current transverse to the direction of the electric field.          

In this Letter, We first show that such an off-diagonal resistance which is symmetric in the spatial coordinates can be non-zero in a 2D system only if the in-plane rotational symmetry is broken down to $C_2$ or below. This implies that although graphene and WSe$_2$ are $C_3$ symmetric individually, for there to be a non-zero symmetric off-diagonal conductivity, the three-fold rotational symmetry of the graphene layer must be broken down to $C_2$ or below in the heterostructure. Following Ref.~[\onlinecite{Aveek_Bid}], we assume the existence of a non-zero uni-axial strain in the graphene layer induced by lattice mismatch, and ascribe the rotational symmetry breaking in the graphene layer to this effect. 
Even so, an important question remains regarding how it is possible for a small amount of strain ($\sim 0.15-0.20 \%$) to produce an almost $20\%$ off-diagonal resistance ($\frac{R_{xy}}{R_{xx}} \sim 20\%$).
To answer this we first show that, interestingly, the transport in a Moire heterostructure can be strongly affected by momentum matching in the extended Brillouin zone rather than in the original Brillouin zone. This can be manifested, among other things, e.g., in the values of the magic angles, also in transport anisotropy. Specifically, such a momentum matching constraint can amplify  a small amount of strain ($\sim 0.15-0.20 \%$) to produce an almost $20\%$ off-diagonal resistance ($\frac{R_{xy}}{R_{xx}} \sim 20\%$).   We can thus show that such amplification of the strain effect leading to an exaggerated transverse resistance response is possible in the experiments performed on the Moire system. We note that the momentum matching required for this constraint occurs at a specific angle, similar to the magic angle twisted bilayer graphene \cite{MacDonald}. However, the system considered here would be an example where the two layers have different lattice constants and the magic angle is a specific value that is a large fraction of $\pi$. By contrast, the magic angles discussed so far in the literature are a parametrically small fraction of $\pi$. In addition to being directly relevant to the experiments in Ref.~[\onlinecite{Aveek_Bid}], our results are important for
detecting anisotropic transport in a nematic system, and may be relevant for detecting off-diagonal anomalous resistance in Kagome charge density waves \cite{Li}. Our work also predicts a magic angle, $\theta\sim 27^0$, in a Moire system which is a significant fraction of $\pi$, and can be experimentally verified in graphene-WSe$_2$ heterostructure.

\textit{Symmetric off-diagonal conductance and rotational symmetry breaking:}
A symmetric off-diagonal resistance, $R_{xy}=R_{yx}\neq 0$ has been observed in the experiments \cite{Aveek_Bid} even without time-reversal symmetry breaking and has been found to persist up to room temperatures. This is surprising given that both graphene and WSe$_2$ individually are $C_3$ rotation symmetric, and hence, as shown below, should exhibit $R_{xy}=R_{yx} = 0$. A non-zero antisymmetric part of the resistance matrix, $R_{xy}=-R_{yx}$, is a Hall resistance that is known to be rotationally symmetric in the 2D plane and requires broken time-reversal symmetry. We start by proving that any system with $R_{xy}=R_{yx}\neq 0$ cannot have any rotation symmetry $C_{n>2}$. To prove this, let us proceed by contradiction and assume that $R_{xy}=R_{yx}\neq 0$. Then we consider the $2\times 2$ symmetrized resistance matrix $R^{(s)}_{ab}$ where $a,b=x,y$. This matrix is real and symmetric and therefore can be diagonalized by an orthogonal matrix (i.e. with columns giving appropriate choices of the orthonormal axes). Furthermore, the eigenvalues of $R^{(s)}$ must be distinct, otherwise $R^{(s)}\propto \mathbf{1}_{2\times 2}$ i.e., the identity matrix. This is because the only symmetric matrix with identical eigenvalues is the $2\times 2$ identity matrix, whose off-diagonal components vanish: $R^{(s)}_{xy}=0$. 
Assuming non-zero off-diagonal resistance ($R^{(s)}_{xy}\neq 0$) then leads to unequal eigenvalues of the symmetric resistance matrix. We can then define $\mathbf{n}_{>}$ to be the direction of the eigenvector of $R^{(s)}_{ab}$ with the higher eigenvalue. This is a specific and unique direction in the material (apart from a sign), $\pm\mathbf{n}_{>}$, along which the resistance is the largest. More specifically, the resistance along any direction $\mathbf{n}$ can be written as $R(\mathbf{n})=\mathbf{n}^T R^{(s)}\mathbf{n}$. Elementary linear algebra can be used to prove that $R(\pm \mathbf{n}_>)>R(\mathbf{n})$ for any $\mathbf{n}\neq \pm \mathbf{n}_>$. This unique direction $\mathbf{n}_>$ breaks any rotational symmetry $C_{n>2}$ down to $C_2$. Thus, any system showing $R_{xy}^{(s)}\neq 0$ must show nematic transport, i.e., different conductances in different directions.  

It should also be noted that $R_{xy}^{(s)}\neq 0$ might be forbidden by symmetries even when $C_2$ is the only rotation symmetry left or there is no rotation symmetry at all. For example, if the $x$ or $y$ direction happens to be along the diagonal axes of $R^{(s)}$, then $R_{xy}^{(s)}=0$. This may also be enforced by a mirror symmetry along $x$ or $y$. Interestingly, the off-diagonal component of the resistance can be conjectured to be a more efficient way to detect anisotropic 
transport. The typical measurement of anisotropic resistance i.e. $R_{xx}\neq R_{yy}$ might be confused with differences in contact resistances between 
the $x$ and $y$ directions and is thus not a smoking-gun evidence of nematicity. This issue would not arise if the conductance is measured off-axis so that $R_{xy}^{(s)}\neq 0$, which is a more robust indicator of nematic order.

\textit{Amplification of rotation symmetry breaking in strained Moire systems:} If the $C_3$ rotational symmetry of graphene is unbroken, then at the Dirac points $v_x=v_y$, precluding non-zero symmetric off-diagonal resistance $R^s_{xy}$. To see why $v_x$ must equal $v_y$ in the presence of $C_3$ symmetry, we can prove it by contradiction by first assuming $v_x \neq v_y$ in the presence of $C_3$ symmetry. The magnitude of velocity at an angle $\frac{2\pi}{3}$ from the $x$-axis is then $\sqrt{\frac{v_x^2}{4}+\frac{3v_y^2}{4}}$, which is less than $|v_x|$ because $v_x > v_y$. But from $C_3$ symmetry, we expect the magnitude of the velocity at an angle $\frac{2\pi}{3}$ from the $x$-axis to be equal to $|v_x|$. It then follows that with unbroken $C_3$ symmetry $V_x$ must be equal to $v_y$. 

One simple way to make $v_x \neq v_y$ and obtain an off-diagonal or anisotropic transport is to apply uniaxial strain to an isotropic system such as graphene. Such strain is known to change the Dirac velocity of graphene \cite{Leyva} and make it anisotropic. This is known to lead to anisotropic conductance simply by making the Fermi velocity anisotropic. However, at first glance, the resistance anisotropy that may follow from strain is expected to be a sub-$1\%$ effect because the magnitude of strain in the experiments of Ref.~[\onlinecite{Aveek_Bid}] is of that order. The graphene-WSe$_2$ Moire heterostructure used in the experiments, on the other hand, shows a symmetric transverse resistance of almost $20\%$, $\frac{R_{xy}}{R_{xx}}\sim 20\%$.  
%Moire systems seem to show a much stronger effect by orders of magnitude. 
Below we explain how this can come about in a Moire system. 

Suppose $G_{1,2}$ are the reciprocal lattice vectors for WSe$_2$. We assume these are normalized so that $|G_{i=1,2}|^2=1$ and $G_1\cdot G_2=1/2$. In these coordinates the $M$ point is at $(1/2,0)$ and $(0,1/2)$. The $K$ point is at $(1/3,1/3)$. The Fourier components of Bloch wave-functions at the $K$ point form a lattice $K_{m,n}=(m+1/3,n+1/3)$  in the extended Brillouin zone. The graphene layer tunnel coupled to WSe$_2$ has the same structure of reciprocal lattice vectors $G'_{i=1,2}$ and Fourier components $K'_{m',n'}$ of the $K$ point wave-functions, except that the magnitudes are different from $K_{m,n}$ because of the difference in the lattice parameters. 
For a Moire system such as the heterostructure of Graphene and WSe$_2$ with distinct lattice parameters, one needs to consider tunneling between states at Bloch momenta $k$ in graphene and $k'$ in WSe$_2$. The tunneling Hamiltonian can be written as a sum over different momentum 
components as,
\begin{align}
H_t=\sum_{m,n}^{m',n'}A(k+K_{m,n},k'+K'_{m',n'})\delta_{k+K_{m,n}-k'-K'_{m',n'}},    \label{eqMoiretunneling}
\end{align}
where the $\delta-$function guarantees the conservation of in-plane momentum by the tunneling process and the momentum-dependent matrix elements $A$ vanish when $k$ or $k'$ are outside the Brillouin zone. Note that, in general, the tunneling does not conserve the Bloch momentum and $k\neq k'$, although it conserves the physical momentum in the 2D extended Brillouin zone. However,
%for WSe$_2$ and graphene, where the reciprocal lattice vectors have a difference in scale,
one needs to find components $K_{m,n}$ and $K'_{m',n'}$ such that $k'-k=K_{m,n}-K'_{m',n'}$ is small. This is because the low-energy states of interest, i.e., those either near the Dirac point of Graphene or the band extrema of WSe$_2$, occur at small momentum $k,k'\sim 0$ relative to the Brillouin zone. For most Moire systems the three momenta, $K_{m,n}-K'_{m'n'}=k-k'$ related by $C_3$ symmetry, dominate the tunneling Hamiltonian. This leads to three vectors $k-k'=\delta G_{i=1,2,3}$ which form the primitive Moire reciprocal lattice vectors.~\cite{MacDonald}

In general, the matching of a pair of wave-function Fourier components $K_{m,n}$ in WSe$_2$ and $K'_{m',n'}$ in graphene at a reasonably small wave vector requires fine-tuning the angle $\theta$ between the lattices of WSe$_2$ and graphene. The magnitudes $|K_{mn}|$ are invariant under the changes of the angle $\theta$. Therefore, we first look for possible matches of the Fourier components by comparing the magnitudes so that $|K_{m,n}|\sim |K'_{m',n'}|$ for some $m,n, m',n'$. Once such a match of the magnitudes is found, we assume that the angle $\theta$ can be chosen to be the angle between these vectors so that we can ensure that $K_{m,n}\sim K'_{m',n'}$ both in magnitude and direction.
The square of the magnitude of the momenta $K_{m,n}$ is given by, 
\begin{align}
&|K_{m,n}|^2=(m+1/3)^2+(n+1/3)^2+(m+1/3)(n+1/3)
\end{align} 
where we have used the fact that $G_1\cdot G_2=1/2$. The scale of $G$ for graphene relative to WSe$_2$ is $|G'_1|/|G_1|=3.32/2.46=1.35$. 
%To get maximal hybridization, we would need one the K points of graphene and WSe2 to nearly match for some $(m,n)$ in Graphene and $(m',n')$ in WSe2.
The magnitudes of the $K$ vectors, in units of the reciprocal lattice vectors of WSe$_2$ (i.e. $|G_1|=1$), are $|K_{(m,n)=(-2,-1)}|=2.0817$ in WSe$_2$  and $|K'_{(m',n')=(-2,0)}|=2.0622$ in graphene. Note that these Fourier components match within $1\%$ of each other.
The matching of these two K points would require the lattices to be rotated very close to an angle 
\begin{align}
    \theta=\cos^{-1}{\frac{K_{m,n}\cdot K_{m'n'}}{\sqrt{K_{m,n}^2 K_{m'n'}^2}}}\sim 27\,\textrm{degrees}.
\end{align}
Based on this model the $WSe_2$-graphene bilayer would show anomalous anisotropic transport only when the relative angle happens to be near this unusually large "magic" angle.

 $C_3$ symmetry implies that these Fourier components of $K$ point wave function in graphene are not the only components that closely match Fourier components 
 in WSe$_2$. Instead, there are two more sets of Fourier components $C_3 K_{-2,0}$ and $(C_3)^2K_{-2,0}$ that are close to the corresponding $K'$ points in WSe$_2$.
Any transport property would still continue to retain $C_3$ rotation symmetry. However, strain dramatically breaks the $C_3$ symmetry by changing the magnitude of the Fourier components so that $|C_3 K_{m,n}|\neq |K_{m,n}|$. 
Since the mismatch between the relevant unstrained reciprocal lattice is about $1\%$, even a small relative strain of magnitude, say $0.3\%$, can lead to a $30\%$ difference in the separation between the $K$ and $K'$ points. This amplified breaking of the $C_3$ symmetry can be large enough to ensure that the $C_3$ rotated Fourier components $C_3 K_{m,n}$ and $C_3 K'_{m'n'}$ become mismatched following the strain and therefore do not contribute to Eq.~\ref{eqMoiretunneling} at small $k,k'$. In what follows,  we will assume that the strain is large enough so that $K_{-2,-1}\sim K'_{-2,0}$ are the only matching Fourier components 
contributing to the $H_t$ in Eq.~\ref{eqMoiretunneling}. This would occur for strains that are large compared to the mismatch of the K-ponits $|\delta G_{i=1,2}|/|G_{i=1,2}|$ relative to the reciprocal latice vectors. In this limit, transport properties of the system would become unconstrained by $C_3$ symmetry, since only one component is present.

\textit{Model calculation for WSe$_2$ - graphene heterostructure}
In this part, we will consider a simplified model where we assume that it is the conduction band of WSe$_2$ that aligns with the Dirac point of graphene so that the spin-orbit splitting in the WSe$_2$ band can be ignored. The motivation  for ignoring the spin-orbit coupling is that the experimental measurement survives at room temperature i.e. a temperature of $T\sim 25 meV$.
The experiment \cite{Aveek_Bid} observes a pair of peaks in $R_{xx}$ as a function of the applied electric field $D$. The linear dependence on density $n$ versus $D$ of edge of the high resistance region suggests that the electronic states of both layers are likely to be involved in the longitudinal transport. This is because the energy difference between the two layers is what is affected by $D$. One of the two features in the longitudinal conductance is expected to be the Dirac point of graphene where the density of states goes to zero. We posit that the other peak in $R_{xx}$ arises from a gap created by the hybridization of a WSe$_2$ band and the Dirac cone. However, it should be noted that this interpretation of the longitudinal conductance measurements is only qualitative and used to motivate the model for WSe$_2$ - graphene heterostructure chosen below, which includes only the conduction band bottom of WSe$_2$. 

The model for WSe$_2$ - graphene that we work with, which includes the conduction band bottom of WSe$_2$
 and 
ignores the induced spin-orbit coupling in  WSe$_2$ and graphene (relative to the tunnel splitting) is written as 
\begin{align}
&H=H_{Graphene}+H_{WSe}+H_t    
\end{align}
 where 
\begin{align}
&H_{Graphene}=(\sigma_x k_x+\sigma_y k_y)P_g\nonumber\\    
&H_{WSe}=((k_x-a_x)^2+(k_y-a_y)^2-\Delta_{off})P_{W}\nonumber\\
&H_t=t(P_t\langle\sigma_z=+1|+h.c).    \label{eq:H}
\end{align}
Here $P_g$, $P_W$ are projections into graphene and WSe$_2$ respectively and $P_t$ is the tunneling from graphene to WSe$_2$. 
The Pauli matrices $\sigma_{x,y,z}$ represent the graphene sub-lattice. The Hamiltonian of WSe$_2$ is chosen to be the low-energy Hamiltonian of the conduction band, which is assumed to be an energy $\Delta_{off}$ above the Dirac point of graphene. Motivated by the presence of two peaks in the longitudinal conductance, the band alignment is assumed to be such that the Dirac point of graphene is closest to the conduction band of WSe$_2$, which has a small spin splitting compared to the valence band. For simplicity, we are also assuming that the WSe$_2$ conduction band only tunnels to the sublattice polarized state $\sigma_z=+1$ in graphene.

The above Hamiltonian represents a single valley. The Hamiltonian in the other valley is  a time-reversed copy of the current Hamiltonian and will contribute in an identical way to computed results.  
As discussed in the previous section, we consider a tunneling model between graphene and WSe$_2$, where the tunneling occurs through a momentum component that is in the vicinity of $K_{2,0}$ in graphene and $K'_{-2,-1}$ in WSe$_2$. This means that the conduction band minimum in WSe$_2$ is offset from the Dirac point of graphene by 
\begin{align}
&(a_x,a_y)=K'_{-2,-1}-K_{2,0}=\delta K.
\end{align}
In the absence of strain breaking rotational symmetry there would have been two other values of $a$ i.e $C_3\delta K$ and $C_3^2\delta K$ that would need to be included in the model. Note that while this model is superficially similar to the Bistritzer-MacDonald model for twisted bilayer graphene (TBLG) Moire system,~\cite{MacDonald} $\delta K$ here is not the difference between the $K$ and $K'$ points in the Brillouin zone but rather the extended Brillouin zone.
   %A $\sigma_z$ rotation can be used to set $k_y=0$ and $k_x=|k|$ in the Graphene Hamiltonian. This generates a phase in the tunneling term, which can be eliminated by a $U(1)$ phase rotation of the $H_{WSe2}$.  

\begin{figure}[t]
\centering
{\includegraphics[width = 1.1\linewidth]{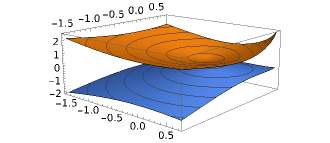}}
\caption{Shows the anisotropy of the band structure induced by the proximity of strained WSe$_2$, with $a_x=a_y=73\mu m^{-1}$, $\Delta_{Off}=1.65$ meV and tunneling $t=4.2 meV$. 
The energy axis (vertical) has a scale of $60$ meV and the momentum axis (horizontal) has a scale of $29 \mu m^{-1}$.  
The parameters are 
%in these units are so that the Dirac velocity of Graphene is chosen to be $1$ and the interlayer tunneling energy $t=1$, which are 
chosen to qualitatively represent the WSe$_2$ and graphene bandstructure. 
}
\label{fig:Bandstructure}
\end{figure}

In the limit $\epsilon=((k_x-a_x)^2+(k_y-a_y)^2-\Delta_{off})$ far from the Dirac point, the low energy effective Hamiltonian can be projected onto the two Dirac point states with an effective Hamiltonian,
\begin{align}
&H_{eff,D}=(\sigma_x k_x+\sigma_y k_y)+(\sigma_z+1)t^2/\epsilon.\label{eq;Heff}
\end{align} 
The energy eigenvalues are $E_{\pm}(k_x,k_y)=t^2/\epsilon\pm \sqrt{k_x^2+k_y^2+(t^2/\epsilon)^2}$. The resulting dispersion (shown in Fig.~\ref{fig:Bandstructure}) breaks rotational symmetry and has a gap at the Dirac point.  This can lead to a peak in the resistivity. In addition, the rotational symmetry breaking arises from the non-zero $(a_x,a_y)$, which enters the expression for $H_{eff,D}$ through $\epsilon$.
The values of the parameters used for this simulation are provided in the caption.
%The resulting dispersion is shown in Fig.~\ref{fig:Bandstructure}.
 Note that the Hamiltonian parameters correspond to fixing the Dirac velocity of graphene and $a_{x,y}$.

The conductance tensor from Boltzmann equation is given by 
\begin{equation}
    \sigma_{ab}=e^2\tau\sum_{n=p,m} \int \frac{d^2k}{(2\pi)^2}v_{n,a} v_{n,b} \sech^2{\beta E_n/4},\label{eq:sigma}
\end{equation}
where $E_{n=\pm}(k)$ is the energy dispersion of the bands, $v_{n,a}=\partial_{k_a}E_n$ and $\tau$ is a scattering time.
\begin{figure}[t]
\centering
{\includegraphics[width = 1.0\linewidth]{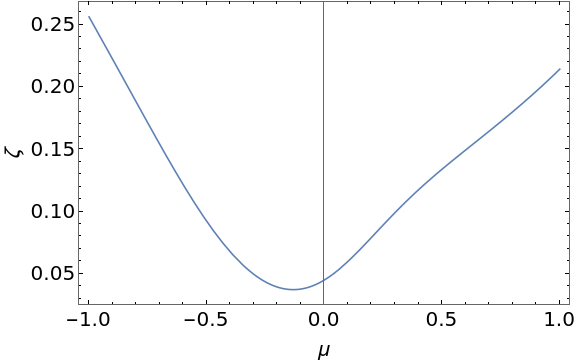}}
\caption{Shows the anisotropy of the conductance i.e. $\zeta=(\sigma_{xy}+\sigma_{yx})/(\sigma_{xx}+\sigma_{yy})$ for the parameters in Fig.~\ref{fig:Bandstructure} as a function of chemical potential $\mu$ for temperature $\beta^{-1}=0.125$ in the dispersion in Fig.~\ref{fig:Bandstructure}. %Despite significant variation we do not see a sign change for the specific model we consider.
}
\label{fig:Conductance}
\end{figure}
The relative off-diagonal conductance $\zeta=(\sigma_{xy}+\sigma_{yx})/(\sigma_{xx}+\sigma_{yy})$, which is computed from the conductance tensor for the 
Hamiltonian Eq.~\ref{eq;Heff} for parameters corresponding to the bandstructure in Fig.~\ref{fig:Bandstructure} is plotted as a function of the 
chemical potential $\mu$ in Fig.~\ref{fig:Conductance}. Note that while the Hamiltonian in Eqn.~\ref{eq:H} is written in units where the Graphene Dirac velocity and the conduction band mass ~\cite{Liu} of WSe$_2$ are set to $1$, which are different from the results in Fig.~\ref{fig:Bandstructure} and Fig.~\ref{fig:Conductance} are presented. As discussed in the introduction, the relative off-diagonal conductance $\zeta$ is a bound on rotational symmetry breaking in the transport and will be referred to as the transport anisotropy.
Note that the degree of rotational symmetry breaking, which is generated by the interaction between the bandstructures of graphene and WSe$_2$, is particle-hole asymmetric in graphene because the WSe$_2$ layer has a particle-hole asymmetric band structure. Specifically, in our example, the bandstructure of WSe$_2$ is represented by a conduction band, which is present at an energy above the Dirac point in graphene without a corresponding state below the Dirac point. %However, the particle-hole asymmetry for our model of the system is not strong enough to switch signs as a function of chemical potential. 
The model we consider for WSe$_2$-graphene is not quantitatively accurate and is only meant to show how strain asymmetries can amplify rotational symmetry breaking and particle-hole asymmetry effects in transport, leading to a quite exaggerated transverse resistance response. Unlike the non-dissipative Hall conductivity which robustly changes sign depending on the sign of the charge carriers, the sign of the symmetric off-diagonal conductivity is non-universal. The switch in the sign of the off-diagonal conductivity seen in the experiments as the chemical potential goes through the Dirac point of graphene is a non-universal effect and will depend on details such as the extent of the particle-hole symmetry breaking in the heterostructure. 

It is important to note that the transport anisotropy $\zeta$ 
in our current computation does not directly arise from the strain in the system. In fact, as discussed in the previous section, we have assumed that the strain was large enough so that we could focus on only one of the three momentum shifts $a=\delta K$ that are related by $C_3$ rotation symmetry are 
included in our model. Thus, there is no reason why the transport anisotropy $\frac{R_{xy}}{R_{xx}}$ should be small if the strain is small.
%This means that the strain dependence of $\zeta$ is maximized in our model. 
Instead, the rotation symmetry breaking in the 
effective Hamiltonian in Eq.~\ref{eq;Heff}, and in turn the value of the transport anisotropy, arises from the momentum shift $a_{x,y}$ of the Dirac points. The anisotropy of the velocity in the dispersion in Fig.~\ref{fig:Bandstructure} occurs at a momentum $k=(a_x,a_y)$, which would vanish as $a_{x,y}$ vanished. It is easy to check that setting $a_x=a_y=0$ restores the rotational symmetry of the effective Hamiltonian in Eq.~\ref{eq;Heff} in the $k_x-k_y$ plane. The magnitude and particle-hole asymmetry of the $\zeta$ depend on all the parameters $t,\Delta_{off}, a_{x,y}$ of our model and is completely non-universal. This proves that the magnitude of the transport anisotropy, specifically, the relative magnitude of the symmetric off-diagonal resistance $\frac{R_{xy}}{R_{xx}}$, does not have to be small if the strain is small, and having been given by non-universal model parameters can easily be $\sim 20\%$ as long as the strain is large compared to the relative momentum mismatch $|\delta G_1|/|G_1|$.   

\textit{Conclusion:} In conclusion, in this paper, we show that a symmetric, off-diagonal resistance in response to a longitudinal current can be non-zero in a 2D system only in the presence of rotational symmetry breaking down to $C_2$ symmetry. Although graphene and WSe$_2$ layers are $C_3$ symmetric, and cannot support non-zero off-diagonal resistance individually, their Moire heterostructure can undergo rotational symmetry breaking due to strain induced by lattice mismatch. By analyzing the tunnel coupling of the layers in a Moire structure, we show that even a small amount of strain (less than $\sim 1\%$) can lead to a greatly exaggerated resistance anisotropy $(\frac{R_{xy}}{R_xx})\sim 20\%$, as seen in experiments.\cite{Aveek_Bid} Our results could be useful for detecting uniaxial anisotropy in nematic systems and in unusual charge density wave states in normal states of Kagome superconductors where the rotation symmetry seems to be broken in the electronic ground state from sixfold to twofold symmetry.\cite{Li} An analogous transport anisotropy has recently been discussed in superconducting vortex lattices, which is a very different system~\cite{Penner}. Our work also predicts a valye for a magic angle, $\theta\sim 27^0$, in a Moire system which is a significant fraction of $\pi$, and can be experimentally verified in graphene-WSe$_2$ heterostructure.

\textit{Acknowledgement:} J. S. acknowledges support from the Joint Quantum Institute and Laboratory of Physical Sciences through the Condensed Matter Theory Center at Maryland. S.T. acknowledges support from ARO Grant No: W911NF2210247 and ONR Grant No: N00014-23-1-2061.


\vskip -6mm
\begin{thebibliography}{24}


\bibitem{Qiao}Z. Qiao, S. A. Yang, W. Feng, W.-K. Tse, J. Ding, Y. Yao, J. Wang, Q. Niu,  Phys. Rev. B \textbf{82},
161414 (2010)
\bibitem{Dean} C. Dean, A.F. Young, L. Wang, I. Meric, G.-H. Lee, K. Watanabe, T. Taniguchi, K. Shepard, P. Kim, J. Hone, Solid State Communications \textbf{152} 1275 (2012)
\bibitem{Cao} Yuan Cao, Valla Fatemi, Shiang Fang, Kenji Watanabe, Takashi Taniguchi, Efthimios Kaxiras and Pablo Jarillo-Herrero, Nature \textbf{556}, 43 (2018)
\bibitem{Wakamura} T. Wakamura, F. Reale, P. Palczynski, M. Q. Zhao, A. T. C. Johnson, S. Guéron, C. Mattevi, A. Ouerghi, and H. Bouchiat, Phys. Rev. B \textbf{99}, 245402 (2019)
\bibitem{Tiwari} Priya Tiwari, Mohit Kumar Jat, Adithi Udupa, Deepa S. Narang, Kenji Watanabe, Takashi Taniguchi, Diptiman Sen, and Aveek Bid, npj 2D Materials and Applications \textbf{6}, 68 (2022)
\bibitem{Xiao} D. Xiao, G.-B. Liu, W. Feng, X. Xu. W. Yao  Phys. Rev. Lett.  \textbf{108},
196802 (2012).
\bibitem{Cresti} A. Cresti, B. K. Nikolic, J. H. García, S. Roche, La Rivista del Nuovo Cimento \textbf{39}, 587 (2016).
\bibitem{Mak} K. F. Mak, K. L. McGill, J. Park, P. L. McEuen, Science \textbf{344}, 1489 (2014).
\bibitem{Lee} J. Lee, K. F. Mak, J. Shan, Nature nanotechnology, \textbf{11}, 421 (2016).
\bibitem{Liu1} J. Liu, Z. Ma, J. Gao, X. Dai, Phys. Rev. X \textbf{9},
031021 (2019)
\bibitem{Wallbank} J.R. Wallbank, D. Ghazaryan, A. Misra, Y. Cao, J.S. Tu, B.A. Piot, M. Potemski, S. Pezzini, S. Wiedmann, U. Zeitler, T.L.M. Lane, S.V. Morozov, Mark Greenaway, Laurence Eaves, A.K. Geim, V.I. Fal'ko, K.S. Novoselov, A. Mishchenko Science  \textbf{353}, 575 (2016)
\bibitem{Qiao1} Z. Qiao, S. A. Yang, W. Feng, W.-K. Tse, J. Ding, Y. Yao, J. Wang,  Q. Niu,  Phys. Rev. B  \textbf{82},
161414 (2010)
\bibitem{Shimazaki} Y. Shimazaki, M. Yamamoto, I. V. Borzenets, K. Watanabe, T. Taniguchi, S. Tarucha, 
 Nature Physics  \textbf{11}, 1032 (2015).
\bibitem{Liu}Gui-Bin Liu, Wen-Yu Shan, Yugui Yao, Wang Yao, and Di Xiao. "Three-band tight-binding model for monolayers of group-VIB transition metal dichalcogenides." Physical Review B 88, 085433 (2013).
\bibitem{Sui} M. Sui, G. Chen, L. Ma, W.-Y Shan, D. Tian,  K. Watanabe, T. Taniguchi, X. Jin,
W. Yao, D. Xiao, Y. Zhang, 
Nature Physics  \textbf{11}, 1027 (2015).
\bibitem{Aveek_Bid} Priya Tiwari, Divya Sahani, Atasi Chakraborty, Kamal Das, Kenji Watanabe, Takashi Taniguchi, Amit Agarwal, Aveek Bid, Nano Letters (in press); arXiv: 2301.01912
\bibitem{Buttiker}Philippe Jacquod, Robert S. Whitney, Jonathan Meair, and Markus Buttiker Phys. Rev. B \textbf{86}, 155118 (2012)
\bibitem{Onsager1} L. Onsager, Phys. Rev. \textbf{37}, 405 (1931)
\bibitem{Onsager2} L. Onsager, Phys. Rev. \textbf{38}, 2265 (1931)
\bibitem{Li} Hong Li, He Zhao, Brenden R. Ortiz, Takamori Park, Mengxing Ye, Leon Balents, Ziqiang Wang, Stephen D. Wilson and Ilija Zeljkovic, Nature Physics \textbf{18}, 265–(2022)
\bibitem{Leyva} M. Oliva-Leyva and Gerardo G Naumis, J. Phys.: Condens. Matter \textbf{26} 125302 (2014)
\bibitem{MacDonald} Rafi Bistritzer and Allan H. MacDonald, 
Proc. Nat. Acad. Sciences, 108 (30) 12233 (2011)
\bibitem{Penner}Alexander-Georg Penner, Karsten Flensberg, Leonid I. Glazman, and Felix von Oppen. arXiv preprint arXiv:2306.09500 (2023).

\end{thebibliography}
\end{document}